\documentclass[journal=ancac3, 
manuscript=article]{achemso}

\usepackage{amsmath}
\usepackage{times}
\usepackage{graphicx}

\title{Graphene-based quantum Hall interferometer with self-aligned side gates} 

\author{Lingfei Zhao}   
\affiliation{Department of Physics, Duke University, Durham, NC 27708, USA.}
\email{lz117@duke.edu}
\author{Ethan G. Arnault}   
\affiliation{Department of Physics, Duke University, Durham, NC 27708, USA.}
\author{Trevyn F. Q. Larson} 
\affiliation{Department of Physics, Duke University, Durham, NC 27708, USA.}
\author{Zubair Iftikhar} 
\affiliation{Department of Physics, Duke University, Durham, NC 27708, USA.}
\author{Andrew Seredinski}   
\affiliation{Department of Physics, Duke University, Durham, NC 27708, USA.}
\author{Tate Fleming}   
\affiliation{Department of Physics and Astronomy, Appalachian State University, Boone, NC 28607, USA.}
\author{Kenji Watanabe}
\affiliation{Research Center for Functional Materials, National Institute for Materials Science, 1-1 Namiki, Tsukuba 305-0044, Japan.}
\author{Takashi Taniguchi}
\affiliation{International Center for Materials Nanoarchitectonics, National Institute for Materials Science, 1-1 Namiki, Tsukuba 305-0044, Japan.}
\author{Fran\c cois Amet} 
\affiliation{Department of Physics and Astronomy, Appalachian State University, Boone, NC 28607, USA.}
\author{Gleb Finkelstein}
\affiliation{Department of Physics, Duke University, Durham, NC 27708, USA.}
\email{gleb@phy.duke.edu}

\begin{document}
\onehalfspacing


\begin{abstract} 
The vanishing band gap of graphene has long presented challenges for making high-quality quantum point contacts (QPCs) -- the partially transparent p-n interfaces introduced by conventional split-gates tend to short the QPC. This complication has hindered the fabrication of graphene quantum Hall Fabry-P\'erot interferometers, until recent advances have allowed split-gate QPCs to operate utilizing the highly resistive $\nu=0$ state.
Here, we present a simple recipe to fabricate QPCs by etching a narrow trench in the graphene sheet to separate the conducting channel from self-aligned graphene side gates. We demonstrate operation of the individual QPCs in the quantum Hall regime, and further utilize these QPCs to create and study a quantum Hall interferometer. 
\end{abstract}

\maketitle

Quantum Hall (QH) interferometers have long provided valuable information about the nature of the edge state and excitations in a variety of QH systems~\cite{Stern2010,Carrega2021}. Key elements of these interferometers are the quantum point contacts (QPCs) -- the gate-defined constrictions that backscatter the QH edge channels. In comparison to GaAs, QPCs are challenging to make in graphene because of its vanishing band gap. The development of graphene QPCs mostly focused on making split-gates on top of boron nitride~\cite{PhysRevLett.107.036602,Zimmermann2017}.  
Recently, such QPCs with $\nu=0$ state underneath the split-gates have been used to make Fabry-P\'erot interferometers~\cite{deprez_tunable_2020,ronen_aharonov-bohm_2020}. Ultra-clean QPCs operating in the fractional QH regime have also been demonstrated~\cite{https://doi.org/10.48550/arxiv.2204.10296}. 

{Although QPCs in GaAs systems are most commonly made in a top-gated geometry, there have also been a number of successful designs based on side-gated QPCs defined by etched trenches. The side-gated design offers some advantages, for example allowing one to induce spin-orbit interactions via asymmetric gating. Most recently, the side-gated QPCs have been fabricated in suspended mechanical nanostructures~\cite{Pogosov2022}.  On the flip side, fabricating clean QPCs is more challenging in the side-gated vs. top-gated geometry due to the roughness introduced by reactive ion etching. However, recently, it has been demonstrated that the cleanness of side-gated QPCs  in graphene can be substantially improved using cryo-etching~\cite{Cleric2019} or AFM lithography~\cite{Kun2020,https://doi.org/10.48550/arxiv.2204.10296}. In terms of fabrication difficulty, top-gated QPCs are easier for GaAs but more challenging for graphene in comparison to the side-gated design. Furthermore, the top gates forming the QPC in graphene have a tendency to be leaky in the lateral direction due to the partially transparent PN junctions formed under them, which is particularly relevant at low magnetic fields~\cite{Zimmermann2017,Veyrat2019}. At a high magnetic field, one of the major applications of QPCs is to explore the anyon statistics in fractional QH interferometers. The recently achieved ultra-clean top-gates QPCs revealed an important complication in this direction –- the edge reconstruction effects of fractional QH states are substantial due to the smooth electrostatic confinement in such structures~\cite{https://doi.org/10.48550/arxiv.2204.10296}. Almost at the same time, a STM investigation evidenced the absence of edge reconstruction for hard-wall confinement in graphene~\cite{https://doi.org/10.48550/arxiv.2210.08152}. Since the confinement potential of side-gated QPCs is much sharper, they may have great potential in exploring the transport of fractional QH edges with suppressed edge reconstruction.}

Here, we {explore side-gated graphene} QPCs by etching self-aligned side-gates and constrictions in graphene/hBN stacks. Such side-gate controlled QPCs may present a simpler alternative to their top-gated counterparts~\cite{Zimmermann2017,deprez_tunable_2020,ronen_aharonov-bohm_2020,https://doi.org/10.48550/arxiv.2204.10296}. The design follows earlier works on graphene nanostructures~\cite{Stampfer2011,Bischoff2015,Bischoff2016_2nd} and our recent work, which showed that self-aligned side gates could efficiently tune the QH filling factor near the edge of graphene~\cite{Seredinski2019}. 
We further characterize the Fabry-P\'erot interferometer defined by a pair of QPCs, and find it in the Coulomb interaction dominated regime~\cite{halperin_theory_2011}. The charging effects are also manifested via phase jumps when localized charge states are filled either in the interferometer or nearby graphene gates.
Following the equilibrium single-particle picture~\cite{de_c_chamon_two_1997}, we extract the edge state velocity of the spin-polarized edge state at $\nu=1$ from the energy scale measured via the bias dependence of the interference patterns. The temperature dependence of the interference visibility  is found to be consistent with this energy scale.

\begin{figure}
\includegraphics[width=1\textwidth]{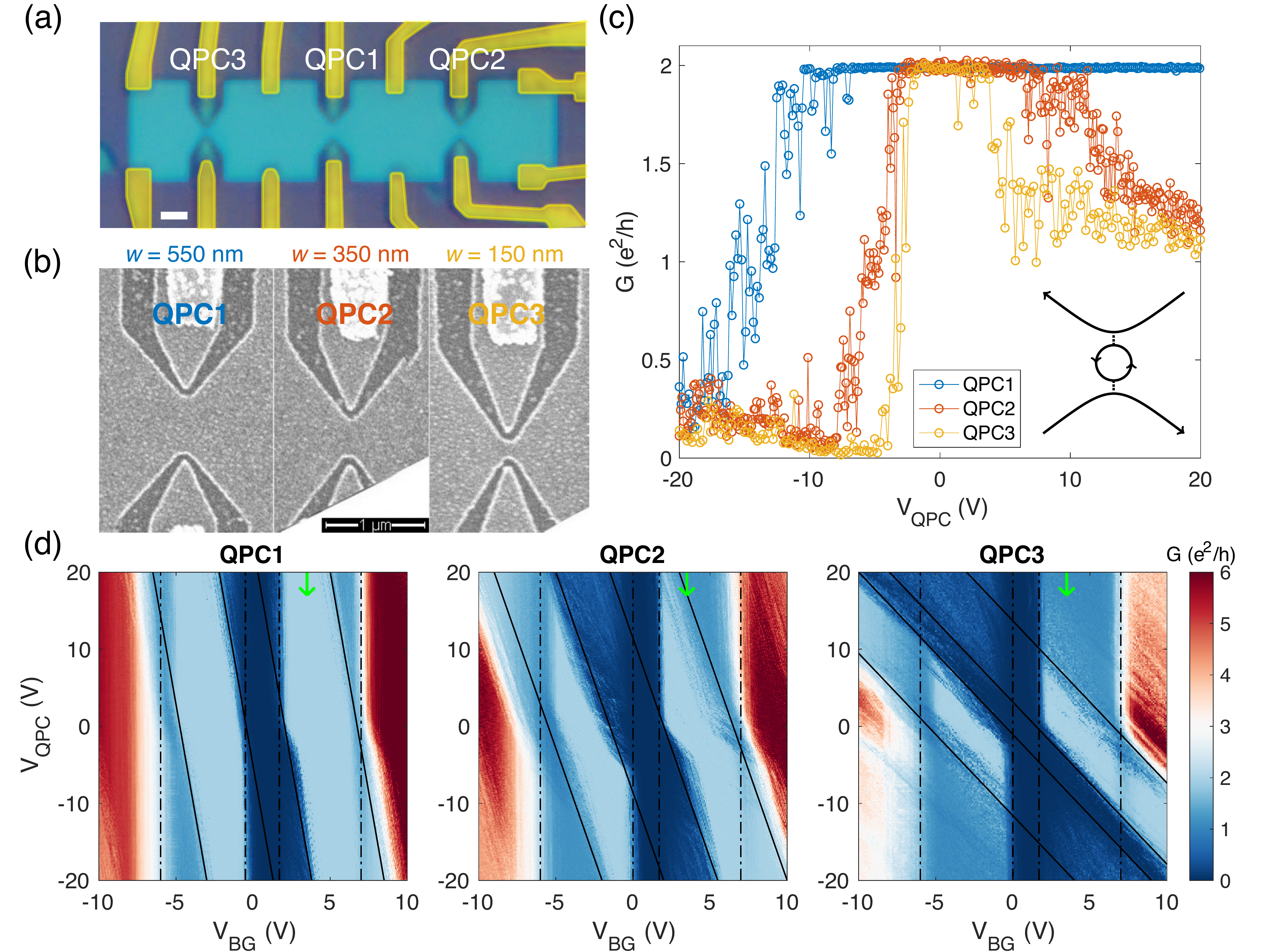}
\caption{\textbf{Characterization of the side-gates controlled graphene QPCs.} 
(a) The false-color optical {(the scale bar is 1 $\mu$m)} and SEM (b) images of the QPCs carved from an encapsulated graphene sheet. The QPCs labeled QPC 1-3 have constriction widths of $w=$ 550, 350 and 150 nm. The regions of graphene separated from the QPC by etching are used as side gates. The smallest distance between the side gates and the constriction is about 50 nm. (c) The two-probe conductance across the QPCs plotted vs. $V_{QPC}$ -- the voltage applied to both side gates which define the QPCs. $B=4$ T, $V_{BG}=3.5$ V, and $\nu=2$, corresponding to the location labeled by {the green arrows} in (d). {For large positive $V_{QPC}$, additional states are introduced in the QPC region (inset), causing equilibration between the counter-propagating channels.} (d) Maps of the QPC conductance for the three QPCs plotted a function of $V_{QPC}$ and $V_{BG}$. {The black sold and dashed lines highlight the filling factor boundaries of the QPC region and bulk respectively.}}
\label{fig1}
\end{figure}

\section{Side-gate controlled QPCs}

Our device consists of a monolayer graphene crystal encapsulated in hexagonal boron nitride~\cite{Dean2010}. The stack is deposited on a SiO$_2$/Si substrate, which serves as the back-gate (BG). We use conventional electron-beam lithography and reactive ion etching to fabricate three QPCs in the same graphene crystal (Fig.~\ref{fig1}(a), see details in Methods). The constriction widths of the three QPCs are $w=$550, 350 and 150 nm, respectively. Both the QPC constrictions and the side gates are formed in the same etching step, as shown in Fig.~\ref{fig1}(b). Here, the QPC channels are horizontal, and the triangular-shaped graphene regions that are etched away from the channel are used as the side gates. Both the side gates and the main region of graphene are contacted via thermally evaporated Cr/Au. The gap between the channel and the side gate at their closest separation is 50 nm and grows wider as one moves further away from the QPC. This design is intended to reduce the undesired gating away from the QPCs.

To characterize the efficiency of the QPCs as a function of constriction width, we measure their two-probe differential conductance, $G$ (Methods). All data presented in this paper are taken at 100 mK unless noted otherwise. In Fig.~\ref{fig1} (c), we show the conductance of all three QPCs as a function of $V_{QPC}$ at a magnetic field of $B=4$ T and a back-gate voltage $V_{BG}=4$ V (bulk filling factor $\nu=2$). When the bulk is n-doped, applying a negative side gate voltage gradually reduces $G$ to zero. This is due to the carrier depletion in the constriction, which causes a reduction of the transmission probability of the edge states. Applying a high positive side gate voltage can introduce additional Landau-levels inside the constrictions, which helps to equilibrate the chemical potential of the counter-propagating edge states~\cite{Zimmermann2017}. Further increasing $V_{QPC}$ results in a reduction of $G$, which saturates at $e^2/h$ when the equilibration process is complete. At this point the constriction effectively serves as a floating contact, thereby doubling the QPC resistance.

As expected, the tuning efficiency of the side gates is higher for the narrower QPCs. Indeed, the extent of the $G\approx 2e^2/h$ plateau in Fig.~\ref{fig1}(c) shrinks as the constriction becomes progressively narrower. While the resulting higher tunability of the QPC is an attractive feature, the degree of conductance quantization eventually degrades. Patches of reduced conductance are seen in the middle of the $\nu=2$ plateau of QPC3. Moreover, the counterpropagating channels rapidly equilibrate with increasing $V_{QPC}$, as clearly visible for QPC3 at $\nu=6$ around $V_{BG}=10$ V in Fig.~\ref{fig1}(d). The degradation is expected due to the smaller separation between the counter-propagating edge states which enables backscattering; the edge disorder introduced by reactive-ion etching~\cite{Bischoff2016} should also have the most effect on the narrowest constriction. The lifting of the quantization becomes even more noticeable for symmetry-broken states which have a smaller energy gap. (These states start to show up at $B= 6$ T in this device and are shown in Supporting Information Fig.~S1.) We therefore conclude that for our design the optimal compromise is obtained for QPC2, which has the medium constriction width of 350 nm.

Fig.~\ref{fig1}(d) shows the conductance maps as a function of the back and side gates in a wide range of bulk filling factors from $\nu= -6$ to $6$ at $B=4$ T. Ideally, QPC conductance should change monotonically between the plateaus~\cite{Zimmermann2017}. However, here multiple resonances appear as lines in the transition regions, corresponding to oscillations in the line cuts presented in Fig.~\ref{fig1} (c). Note that the slope of these resonances in the gate-gate map is negative, apparently tracing lines of constant electron density in the QPC regions. This is an indication that the resonances are caused by the charging of localized states within the constrictions~\cite{Bischoff2015,Bischoff2016}. From the slopes, we roughly estimate the side-gates efficiency in tuning of electron density as 1/10, 1/6 and 1/2 times the back-gate efficiency for QPC1, QPC2 and QPC3 respectively.

\begin{figure}
\includegraphics[width=0.9\textwidth]{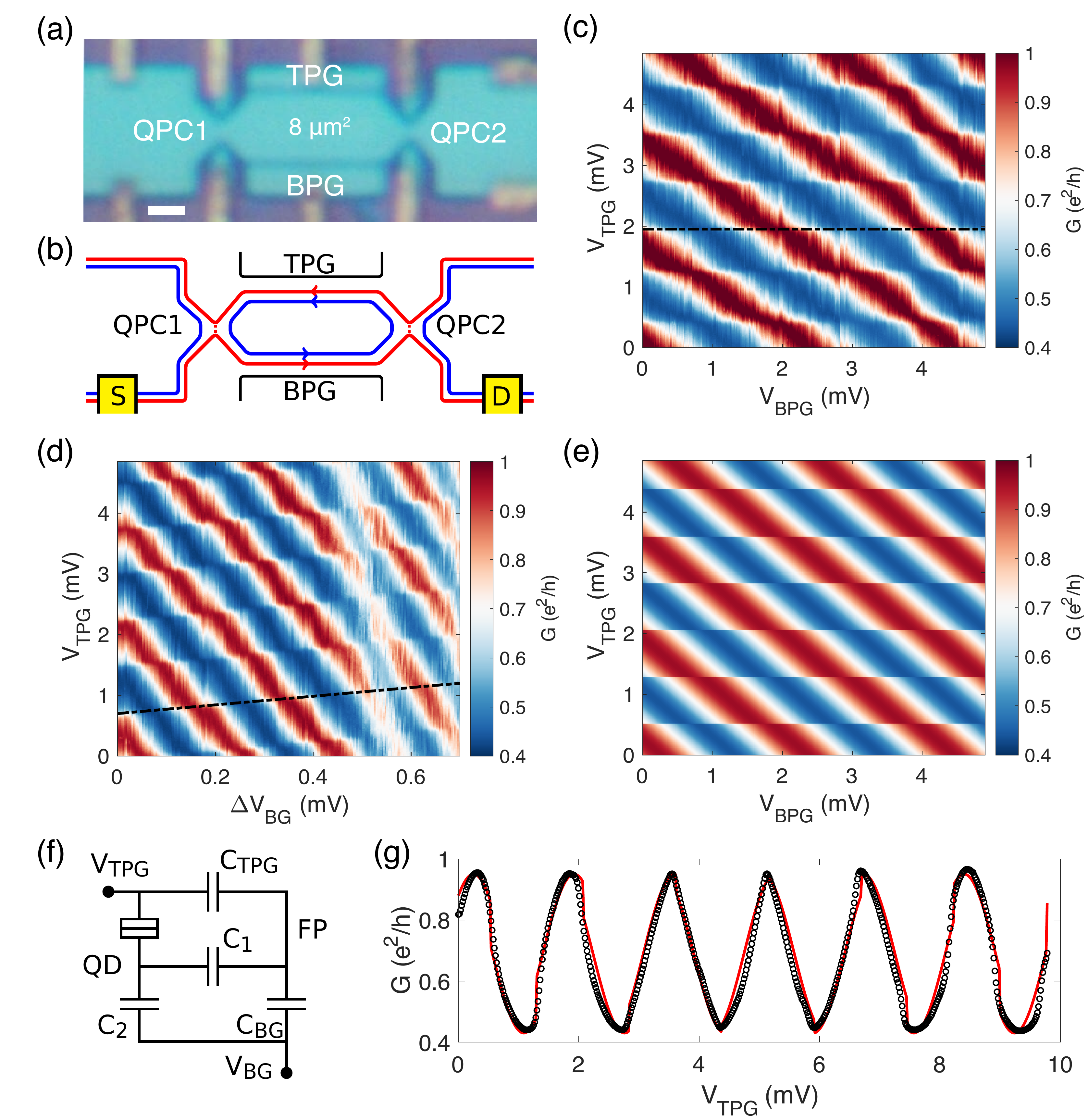}
\caption{\textbf{Phase jumps of the Fabry-P\'erot interferometer.} 
(a) Optical image of the Fabry-P\'erot interferometer. Scale bar is 1 $\mu$m. (b) Sketch showing the propagation of the $\nu=$ 2 edge states. Two distinct edge channels are obtained at $B=$ 6 T and $V_{BG}\approx$ 3.05 V. The QPC voltages are $V_{QPC1}=- 20$ V and $V_{QPC2}=- 8$ V, resulting in the outer channel being partially transmitted while the inner channel is fully reflected. (c,d) Two-probe conductance $G$ across the interferometer measured as a function of c) TPG and BPG and d) TPG and BG. The dashed lines mark the position of the phase jumps. (e) Simulated conductance map corresponding to panel (c) with parameters obtained from the fit in panel (g). (f) The capacitance network model we use to explain the phase jumps observed in (c,d). The unintentional quantum dot (QD) is a part of the top plunger gate (see text). The estimated capacitances are $C_{TPG}\approx77$ aF, $C_{BG}\approx800$ aF, $C_1\approx15$ aF and $C_2\approx130$ aF. (g) A vertical cut of the conductance map in panel (c) at $V_{BPG}=0$ (black dots) agrees well with the fit (red curve) corresponding to the model in panel (f). }
\label{fig2}
\end{figure}

\section{Fabry-P\'erot Interferometer and Graphene Plunger Gates}
 
To further explore the operation of the graphene gates, we designed a Fabry-P\'erot interferometer using QPC1 and QPC2, which both showed good quantization of the $\nu=1$ and 2 states at $B=6$ T (see Supporting Information Fig.~S1). To this end, we etched $\sim100$ nm wide trenches into the graphene region between QPC1 and QPC2, thereby detaching the two middle metal contacts that were used to measure Fig.~\ref{fig1}. The advantage of severing the contacts is two-fold. First, the edge states that used to be equilibrated by these contacts can now preserve their phase coherence in the region between the two QPCs, thereby forming the interferometer, see Fig.~\ref{fig2}(a). Second, the two graphene strips attached to the former contacts now serve as the top and bottom plunger gates (TPG and BPG). Voltages $V_{TPG}$ and $V_{BPG}$ applied to the plunger gates tune the positions of the quantum Hall edge states, which affect the effective area of the cavity between the two QPCs. The resulting interferometer cavity is about 8 $\mu m^2$ in area and 12 $\mu$m in perimeter. 

We first study the $\nu=2$ case, when the interferometer has two distinct edge channels. The back-gate voltage is set at $V_{BG}$ = 3.05 V, corresponding to the beginning of the bulk $\nu=$ 2 plateau, which extends from 2.8 V to 6.8 V. The QPC voltages are fixed at $V_{QPC1}=-$ 20 V and $V_{QPC2}=-$ 8 V. Here, the inner channel is fully reflected while the outer channel is partially transmitted into the cavity and is the one causing the interference pattern, see schematics in Fig.~\ref{fig2}(b). 
With these parameters, the transmission probabilities of the outer channel are estimated to be $\sim0.5$ (see Supporting Information).

We start by applying voltages to both plunger gates; increasing these voltages moves the interfering channel towards the graphene edge, thereby increasing the enclosed area $A$. If the magnetic field $B$ is held constant, a change $\delta A$ in the area modulates the AB phase of electrons traveling around the cavity by $\delta \phi=2 \pi e B \delta A /h$. This modulation can be observed by measuring the differential conductance across the interferometer while tuning both $V_{TPG}$ and $V_{BPG}$ (Fig.~\ref{fig2}(c)). We observe periodic interference fringes in a wide range of gate voltages, with similar periods of about 2 mV, indicating that the two gates have similar efficiency. 

We next focus on the small phase jumps observed in Fig.~\ref{fig2}(c) (one of them is marked by the black dashed line). The features are periodic in TPG, indicating regular charging of some localized states. The features are not affected by the BPG, resulting in their appearance as horizontal lines. Apparently, these states are located in the TPG region, and the BPG is too far away to have a sizable effect. In Fig.~\ref{fig2}(d), we plot the same interference pattern as a function of TPG and BG. The features now acquire a slope vs. $V_{TPG}$ and $V_{BG}$ (see the black dashed line as an example), which indicates that the states are influenced by both grates. In comparison to the AB fringes associated with the interferometer, these phase jumps appear twice more frequently in $V_{TPG}$, but five times less frequently in $V_{BG}$ (see the separation between jumps in Fig.~\ref{fig2}(d)). Most importantly, their slope is positive, opposite to that of the AB fringes. 
 
These observations suggest that the states responsible for the small phase jumps are localized not in the interferometer, but in the region of graphene attached to the TPG, which effectively forms a small quantum dot. The states in this dot are coupled to the gate electrode via tunneling, and applying positive TPG voltage depletes this dot of electrons, in contrast to the same voltage adding electrons in the interferometer. This is the reason why the resulting phase jumps follow lines of a positive slope in the $V_{TPG} - V_{BG}$ plane of Fig.~\ref{fig2}(d). The graphene region connected to TPG is about 4-6 times smaller than the interferometer, in agreement with the observed smaller capacitance of these states to the BG as compared to the states of the interferometer.

Once the origin of the phase jumps has been identified, it is straightforward to account for them by considering the capacitive network shown in Fig.~\ref{fig2}(f). Here, the localized states in the TPG region (QD) are tunnel coupled to $V_{TPG}$. Their capacitances to the interferometer (FP) and BG are $C_1$ and $C_2$ respectively. The capacitances of the interferometer to the TPG and BG are $C_{TPG}$ and $C_{BG}$. The amplitude of the phase jumps is given by $2\pi C_1/(C_1+C_2)$.
We approximate the oscillatory part of conductance by a simple $\delta G \propto \cos(\delta \phi)$ appropriate for our relatively high base temperature of 100 mK where higher-harmonics have been sufficiently suppressed by thermal smearing effect (see the next section). 
In Fig.~\ref{fig2}(g), we plot the conductance vs. TPG (dots), corresponding to the vertical cross-section on Figs.~\ref{fig2}(c). The distortions of the oscillations introduced by the phase jumps are well captured by the fit (red line). The fit gives $C_{TPG}\approx77$ aF and $C_1\approx C_2/9$.
Using the parameters obtained in this fit, we can calculate the expected interference pattern. The resulting conductance pattern is shown in Fig.~\ref{fig2}(e), demonstrating a good qualitative agreement with the experiment (Fig.~\ref{fig2}(c)). From the $V_{BG}$ period of the interference pattern $\sim0.2$ mV and of the phase jumps $\sim1.1$ mV in Fig.~\ref{fig2}(d), we further estimate $C_{BG}\approx800$ aF and $C_1+C_2\approx145$ aF. Therefore, we have $C_1\approx15$ aF and $C_2\approx 130$ aF.
We note that the BPG does not produce similar features, indicating that the metal electrode is better coupled to the nearby graphene region.

\section{Energy dependence of conductance oscillations}

\begin{figure}
\includegraphics[width=1\textwidth]{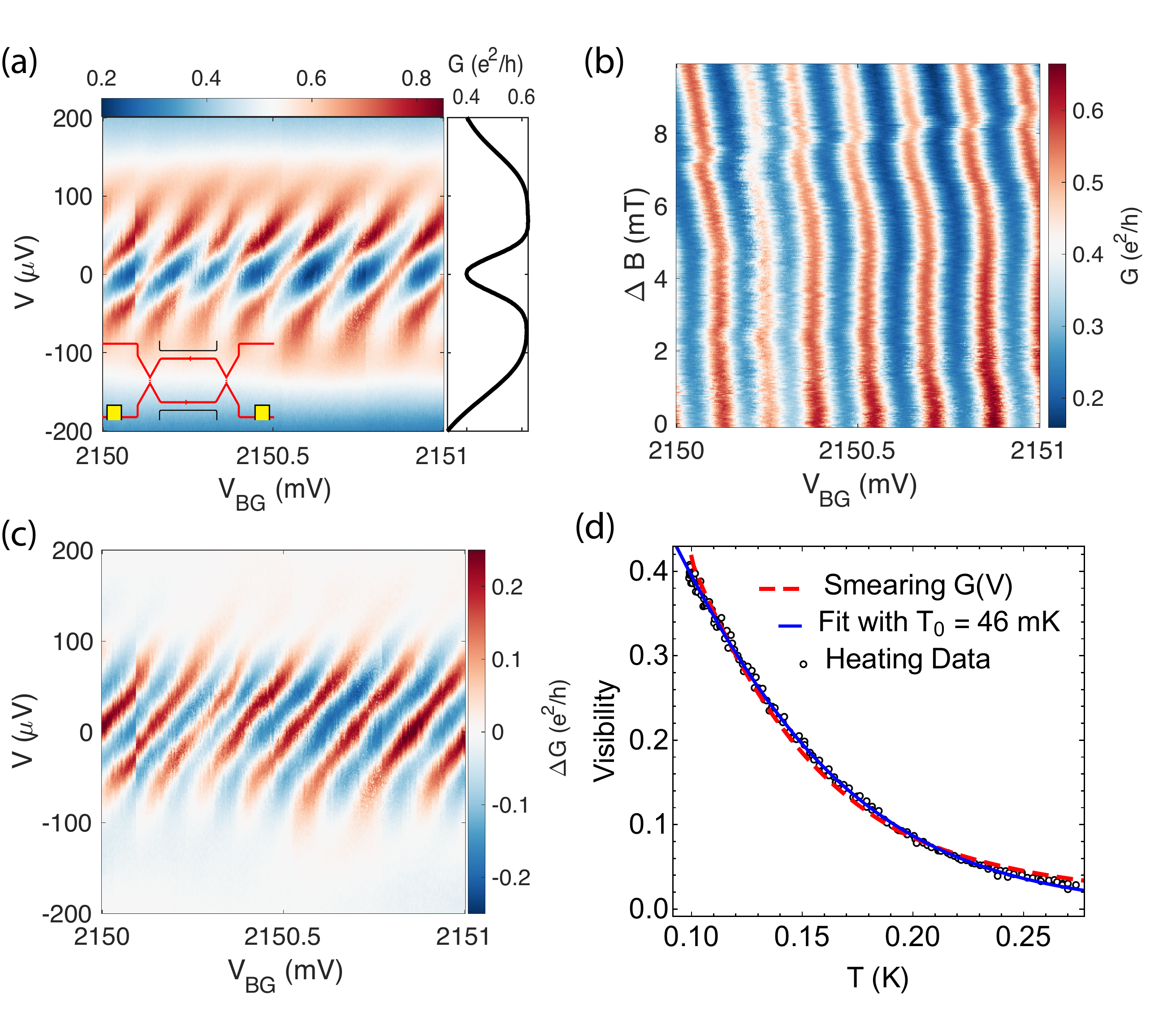}
\caption{\textbf{Energy dependence of AB interference.} 
(a) The interferometer conductance measured as a function of bias $V$ and $V_{BG}$ at $\nu=$ 1 (schematic indicated in the inset). The QPC transmission probabilities are $\mathcal{T}_{QPC1}\sim0.7$ and $\mathcal{T}_{QPC2}\sim0.6$. The right inset shows the conductance averaged over the gate voltage, $\langle G(V) \rangle_{BG}$, plotted as a function of bias. (b) Interference pattern vs. $V_{BG}$ and the incremental magnetic field change $\Delta B$. (c) Conductance of panel (a) with gate-averaged background $\langle G(V) \rangle_{BG}$ (inset of panel a) subtracted to reveal the interference fringes. (d) The visibility of zero-bias oscillation pattern plotted vs. temperature. The blue solid curve is a fit in the form of Eq.~(\ref{eq2}) with $T_0=46$ mK and the red dashed line is obtained from numerically convolving the data in panel (a) with the derivative of the Fermi function.}
\label{fig3}
\end{figure}

We next explore the energy dependence of the interference fringes by tuning the temperature and bias. We chose to set the bulk filling factor at $\nu=1$, when only a single spin-polarized edge channel is present. We plot the conductance pattern as a function of bias $V$ and $V_{BG}$ in Fig.~\ref{fig3}(a). The map shows contours of constant phase, which are sometimes interpreted as non-interacting AB oscillations, although it has been demonstrated that the charging effects are significant~\cite{ofek_role_2010,sivan_interaction-induced_2018} even at $\nu=1$~\cite{roosli_observation_2020}. We estimate (see Supporting Information) that in our case the charging energy exceeds the level spacing. In fact, the familiar Coulomb diamonds can be recovered by integrating the differential conductance to plot the current through the dot (Fig.~S4). The diamonds are strongly asymmetric because the capacitance to the gates dominates the capacitance to the source and drain contacts. The vertical boundaries of the diamonds are not visible in the differential conductance but are revealed in the current map.

The strong role of the Coulomb interactions is confirmed by Fig.~\ref{fig3}(b), which shows the map of conductance in the $V_{TPG} - \Delta B$ plane, where $\Delta B$ is the deviation of magnetic field from 6 T. The AB period of the interferometer should be $\sim 0.5$ mT, and on that scale the interference fringes show negligible slope, which indicates the dominant role of the charging energy~\cite{halperin_theory_2011}. Similar nearly vertical conductance fringes were observed in the other measurement at $\nu=1$ dominated by Coulomb interactions~\cite{roosli_observation_2020}. 

Nonetheless, the description in terms of Fabry-P\'erot interference is still possible. To that end, we average the conductance over the gate voltage, $\langle G(V) \rangle_{BG}$, as shown in the right inset of Fig.~\ref{fig3}(a). We then subtract $\langle G(V) \rangle_{BG}$  from the data in Fig.~\ref{fig3}(a) and obtain a simple stripe pattern of Fig.~\ref{fig3}(c), which strongly resembles the simulated conductance map of resonant tunneling plotted in Fig.~S8. Note that the background subtraction does not affect the amplitude of conductance oscillations vs. $V_{BG}$. 
 
The conductance fringes in Fig.~\ref{fig3}(c) have a positive slope, where increasing $V_{BG}$ increases the AB phase $\delta \phi$, as we have seen in Fig.~\ref{fig2}. On the other hand, increasing $V$ reduces the ``traveling phase'' $\phi_0=(-eV/\hbar v) L$, where $L$ is the perimeter of the interferometer and $v$ is the velocity of the edge states, which is renormalized by interactions. We use Fig.~\ref{fig3}(c) to obtain the bias oscillation period of $V_0\approx$ 80 $\mu$V. From here, we estimate the edge state velocity $v=e V_0 L/h\approx$ 2.3$\times$10$^5$ m/s, which agrees with the previously reported values~\cite{deprez_tunable_2020,ronen_aharonov-bohm_2020}.

The variation of the geometrical phase $\phi_0=\frac{2 \pi E}{eV_0}$ with energy results in smearing of the conductance pattern. We find that in our temperature range $T=100-250$~mK it is sufficient to approximate the conductance oscillations as $\delta G \propto \cos(\delta \phi+\phi_0)$. From here, we obtain the thermal dependence of the oscillations:
\begin{equation}
\delta G\propto \int_{-\infty}^{\infty} \cos(\delta\phi+\frac{2 \pi E}{eV_0})\frac{\partial f(E,T)}{\partial E}dE \propto \frac{T/T_0}{\sinh(T/T_0)} \cos(\delta \phi),
\label{eq2}
\end{equation}
where $f$ is the Fermi distribution function and $T_0=eV_0/2\pi^2k_B$. Therefore, the visibility of the oscillations, defined as $(G_{max}-G_{min})/(G_{max}+G_{min})$, is expected to be proportional to $T/\sinh(T/T_0)$. To extract the visibility from the experimental data, we measure the conductance maximum (minimum) as the average height of several peaks (valleys) in $G(T)$ measurement of Fig.~S5. The resulting Fig.~\ref{fig3}(d) shows the measured zero-bias visibility vs. temperature (black dots) and the fit (blue solid line). Note that 
the expression of Eq.~(\ref{eq2}) does not simply reduce to $\exp(-T/T_0)$. The fit yields $T_0=46$ mK, which closely matches the value $eV_0/2\pi^2k_B\approx 47$ mK with the value of $V_0$ obtained from Fig.~\ref{fig3}(c). 
 
Another way to compare the energy and temperature scales is to numerically convolve the transmission extracted from the $G(V)$ pattern in Fig.~\ref{fig3}(a) with the Fermi-Dirac distribution at elevated temperatures. The result is shown as the red dashed line in Fig.~\ref{fig3}(d), also agreeing with the measured temperature dependence. The fact that both numerical convolution of $G(V)$ and fitting with Eq.~(\ref{eq2}) produce consistent results verifies that thermal broadening is the main origin of the visibility suppression at elevated temperatures. 


In summary, we present a simple recipe to fabricate quantum point contacts with self-aligned side-gates in graphene, which operate in the quantum Hall regime. A reduction in the constriction widths of the QPCs improves their tunability at the expense of the degree of the conductance quantization, suggesting an optimal width in the $300-400$ nm range. We use this technique to define a quantum Hall Fabry-P\'erot interferometer and explore the charging patterns both for $\nu= 1$ and $2$, studying their dependence on gate voltages, bias and temperature. In the future, additional screening layers such as top and bottom graphite gates can be used to reduce the charging effects~\cite{nakamura_aharonovbohm_2019}. Such designs, pioneered in GaAs could enable exploration of the non-Abelian statistics in the fractional quantum Hall regime~\cite{nakamura_direct_2020,PhysRevLett.111.186401}. The quality of the side-gates and QPCs can also be substantially improved with emerging cryoetching~\cite{Cleric2019} and AFM lithography~\cite{Kun2020,https://doi.org/10.48550/arxiv.2204.10296} techniques.

\section{Methods}

\subsection{Sample Fabrication}
The graphene and h-BN crystals are exfoliated onto SiO$_2$ substrates and then assembled together by the conventional dry transfer technique using PC/PDMS stamps. The contacts and trenches are patterned by electron beam lithography using a layer of PMMA resist, developed in cold IPA/DI water mixture (ration 3:1) and then further etched by CHF$_3$/O$_2$ plasma in a reactive ion etcher. The Cr/Au (1 nm/100 nm) metal leads are deposited at 10$^{-7}$ mbar base pressure in a thermal evaporator.  

\subsection{Measurement}
The device is cooled down to the base temperature of $\sim 100$ mK in a Leiden dilution refrigerator wired with resistive coaxial lines and low pass filters. The two-terminal differential resistance of individual QPCs is measured with 1 nA, 15 Hz square-wave excitations with a digital acquisition board and a homemade voltage preamplifier. For a better signal-to-noise ratio, the two-terminal differential conductance of the interferometer is measured with a lock-in amplifier and homemade low noise current preamplifier with a current to voltage gain of 10$^7$. The amplitude of the voltage excitation applied from the lock-in amplifier is 7 $\mu$V and the frequency is 577 Hz. In this case, the device is measured in series with two 5 kOhm filters, whose resistance is then subtracted from the measured resistance. Similarly, the voltage drop across the filters is subtracted from the applied bias to obtain 
the bias across the sample. By calibrating against the quantum Hall resistance of the sample, an overall gain correction of 1.0638 is obtained to account for the RC filters, voltage dividers and buffers used in the system. The DC bias and gate voltages are applied by a combination of NI USB-6363 and DAC8728 controlled by NI USB-6501.

\acknowledgement
We greatly appreciate stimulating discussion with H. U. Baranger.
Transport measurements conducted by L.Z., E.G.A., and T.F.Q.L. were supported by NSF award DMR-2004870. 
Lithographic fabrication and characterization of the samples by L.Z. and A.S., as well as the project guidence by Z.I and G.F were supported by the Division of Materials Sciences and Engineering, Office of Basic Energy Sciences, U.S. Department of Energy, under Award No. DE-SC0002765. 
T.F. and F.A. {were supported by a URC grant at Appalachian State University.}
K.W. and T.T. acknowledge support from the Elemental Strategy Initiative conducted by the MEXT, Japan, (grant no. JPMXP0112101001), JSPS KAKENHI (grant no. JP20H00354) and CREST (no. JPMJCR15F3, JST).
The sample fabrication was performed in part at the Duke University Shared Materials Instrumentation Facility (SMIF), a member of the North Carolina Research Triangle Nanotechnology Network (RTNN), which is supported by the National Science Foundation (Grant ECCS-1542015) as part of the National Nanotechnology Coordinated Infrastructure (NNCI).


\providecommand{\latin}[1]{#1}
\providecommand*\mcitethebibliography{\thebibliography}
\csname @ifundefined\endcsname{endmcitethebibliography}
  {\let\endmcitethebibliography\endthebibliography}{}

\section*{Supporting Information}
\global\long\def\theequation{S\arabic{equation}}
\global\long\def\thefigure{S\arabic{figure}}
\setcounter{equation}{0}
\setcounter{figure}{0}

\section{Characterization of QPC transmissions}

\begin{figure}
\includegraphics[width=1\textwidth]{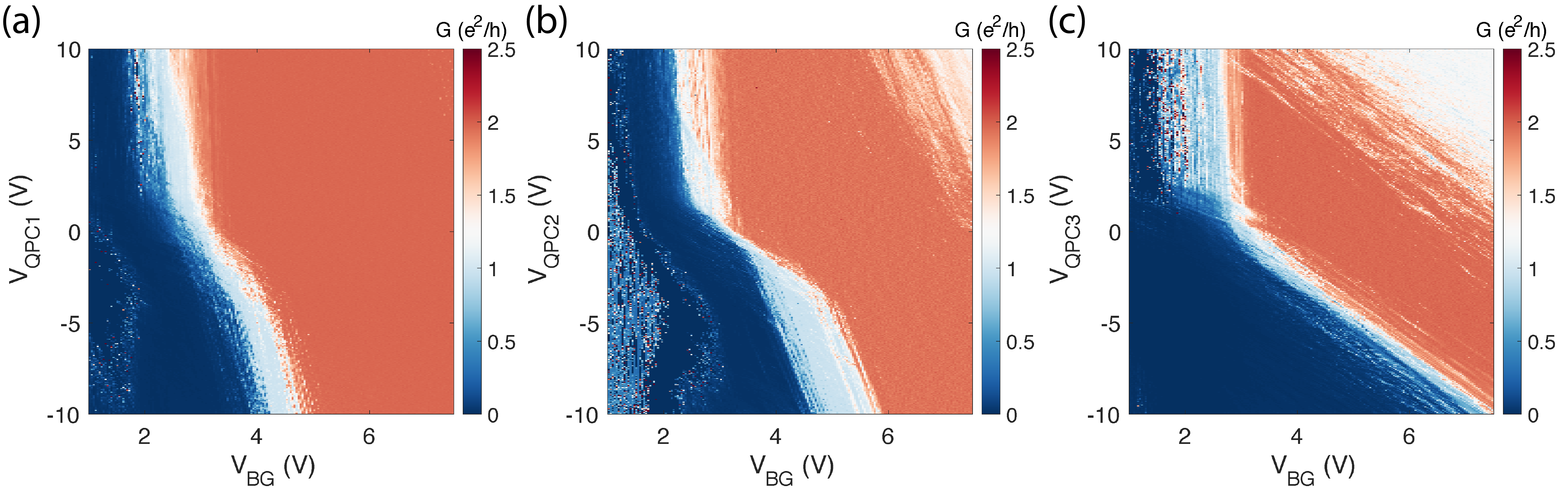}
\caption{Two-probe differential conductance of a) QPC1 and b) QPC2 before the device was etched into an interferometer. The data measured at 6T shows quantized $\nu=1$ regions. In comparison, $\nu=1$ has not formed yet for the narrowest QPC3, shown in (c).}
\label{sup0}
\end{figure}

Before etching the device to form an interferometer, QPC1 and QPC2 are characterized at 6 T showing good quantization for $\nu=1$ (Fig.~\ref{sup0}). Once the sample was etched to form the interferometer, the transmission coefficient of a particular QPC (1 or 2) can be estimated by measuring the two-terminal conductance of the interferometer, $G$, while keeping the other QPC (2 or 1) open. 

\begin{figure}
\includegraphics[width=1\textwidth]{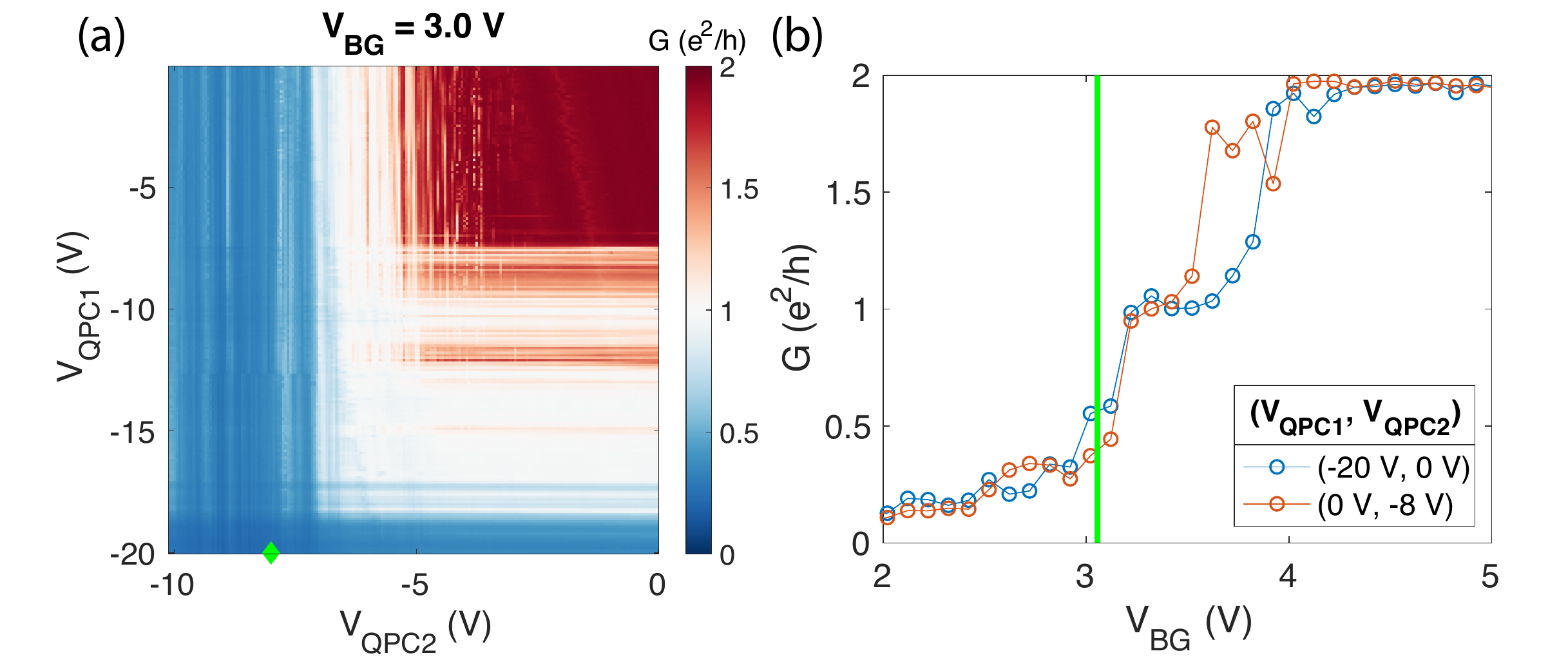}
\caption{\textbf{QPC operation for Fig.~2.} 
(a) The interferometer conductance $G$ as a function of QPC voltages at $V_{BG}=3$ V ($\nu=2$). The green diamond marks the QPC voltges used in Fig.~2 ($V_{QPC1}=-20$ V and $V_{QPC2}=-8$ V.).
(b) $G$ as a function of $V_{BG}$ at $V_{QPC1}=-20$ V, $V_{QPC2}=0$ (blue) V and $V_{QPC1}=0$ V, $V_{QPC2}=-8$ V (red). The vertical green line shows the position of the back-gate voltage in Fig.~2 ($V_{BG}=3.05$ V). From the intersections we determine $\mathcal{T}_{QPC1}\sim0.6$ and $\mathcal{T}_{QPC2}\sim0.4$.}
\label{sup1}
\end{figure}

As shown in Fig.~\ref{sup1}(a), we first measure $G$ as a function of both QPC voltages at $V_{BG}=3$ V ($\nu=2$). The green diamond marker labels the QPC operation point for Fig. 2 ($V_{QPC1}=-20$ V and $V_{QPC2}=-8$ V). At $V_{QPC1,2}=0$ V, both QPCs are open ($G=e^2/h$). From the conductance at $(V_{QPC1},V_{QPC2})=(-20,0)$ V and $(0,-8)$ V, we find the QPC transmissions to be around 0.4 for both QPC1 and QPC2. For the data shown in Fig. 2, we use a slightly higher $V_{BG}=3.05$ V and the QPC transmissions are found to be higher (see Fig.~\ref{sup1}(b) where the green line marks $V_{BG}=3.05$ V). Here, the estimation of QPC transmissions is $\mathcal{T}_{QPC1}\sim0.6$ and $\mathcal{T}_{QPC2}\sim0.4$. 

\begin{figure}
\includegraphics[width=1\textwidth]{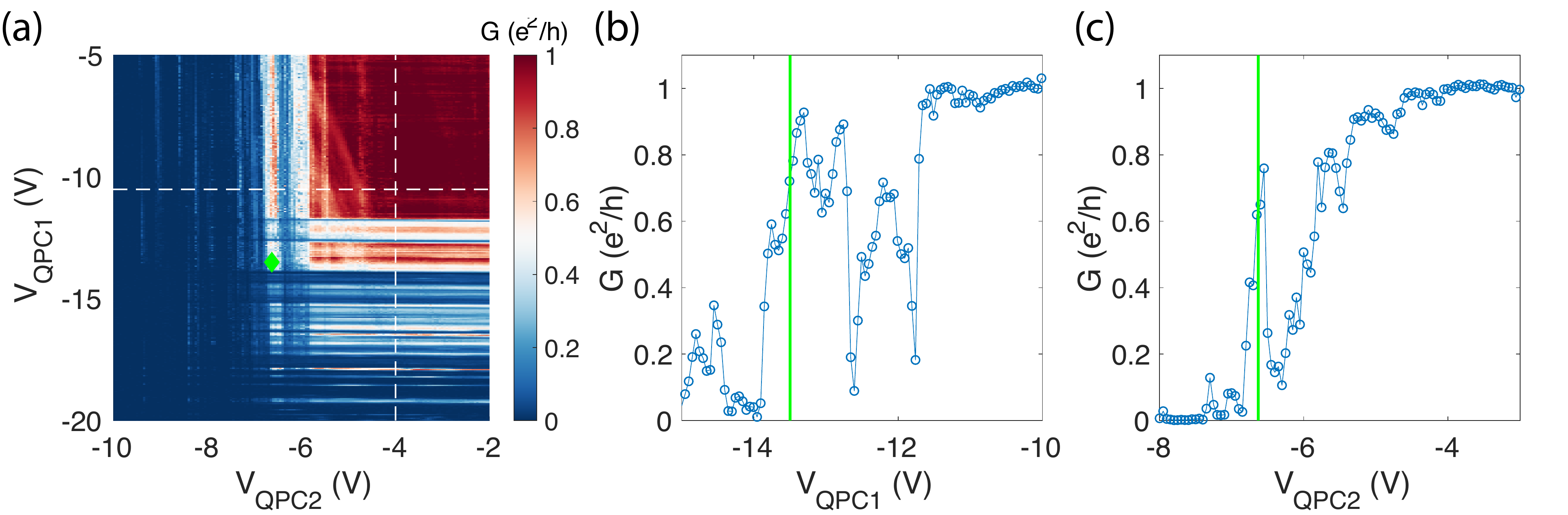}
\caption{\textbf{QPC operation map for Fig.~3.} 
(a) The interferometer conductance $G$ as a function of QPC voltages at $V_{BG}=2.15$ V ($\nu=1$). The green diamond marks the QPC voltges used in Fig.~3 ($V_{QPC1}=-13.5$ V and $V_{QPC2}=-6.63$ V.). 
(b) A cut of (a) along the vertical white dashed line showing $G$ as a function of $V_{QPC1}$ while QPC2 is open. The green line labels QPC1 operation voltage. (c) A cut of (a) along the horizontal white dashed line showing $G$ as a function of $V_{QPC2}$ while QPC1 is open. The green line labels QPC2 operation voltage. From the intersections in (b, c) we determine $\mathcal{T}_{QPC1}\sim0.7$ and $\mathcal{T}_{QPC2}\sim0.6$.}
\label{sup2}
\end{figure}

To determine the QPC transmissions at the $\nu=1$ operating point, in Fig.~\ref{sup2} (a), we plot $G$ as a function of both QPC voltages at the same back-gate voltage used in Fig. 3, i.e. $V_{BG}=2.15$ V. The green diamond marker labels the QPC operation point for Fig. 3 ($V_{QPC1}=-13.5$ V and $V_{QPC2}=-6.63$ V). Compared to $\nu=2$,  the QPC transmissions at $\nu=1$ are much more sensitive to the gate voltages, and the cross-talk between the two QPC gates becomes noticeable. In order to estimate the transmission of QPC1 more accurately in the presence of cross-talk, in Fig.~\ref{sup2}(b) we plot $G$ as a function of $V_{QPC1}$ at $V_{QPC2}=-4$ V, labeled by the vertical white dashed line in panel (a). This value of $V_{QPC2}$ is not too far from the operation point, yet QPC2 is already open. Similarly, for determining transmission of QPC2, in Fig.~\ref{sup2}(c) we plot a cut at $V_{QPC1}=-10.5$ V, corresponding to the horizontal white dashed line in panel (a).The QPC operation gate voltages labeled by green lines in panel (b) and (c), from which we estimate $\mathcal{T}_{QPC1}\sim0.7$ and $\mathcal{T}_{QPC2}\sim0.6$.

Note that between measuring in the $\nu=2$ and $\nu=1$ regimes, the sample condition changed due to a sudden power-off of the electronics controlling the gates. Although the QPCs were still operating properly, the charge neutrality point shifted towards the p-side by $\sim0.5$ V in $V_{BG}$.

\section{Map of current through the interferometer}

\begin{figure}
\centering
\includegraphics[width=0.6\textwidth]{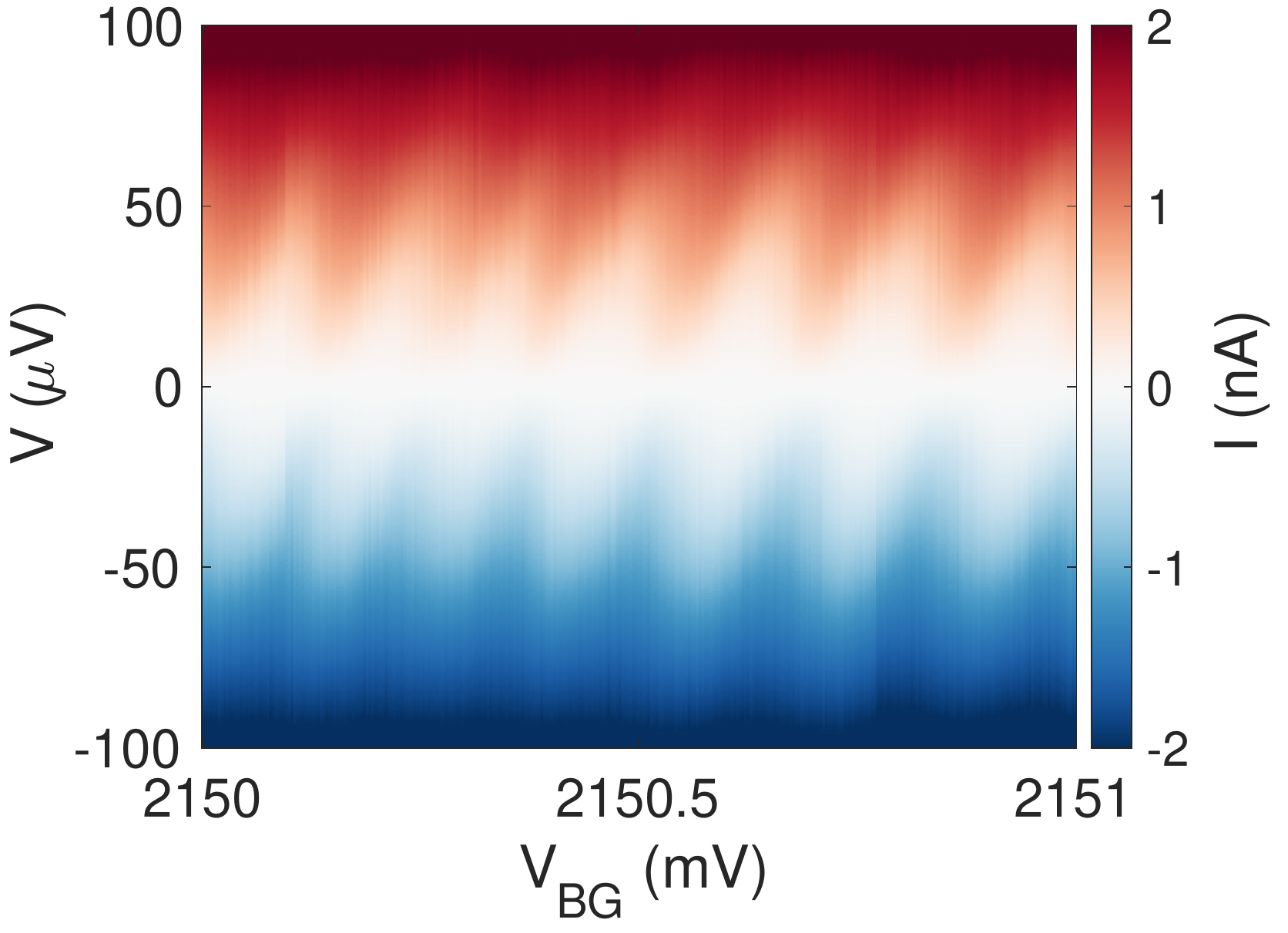}
\caption{The integral of the differential conductance in Fig.~3(a), showing the current through the interferometer as a function of bias and back-gate voltage.}
\label{diamond}
\end{figure}

As demonstrated in Ref.~\cite{S_roosli_observation_2020}, quantum Hall Fabry-P\'erot interferometers have close relation to quantum dots in the quantum Hall regime. When the transmissions across the QPCs are high, the device is operated in the interferometer regime. Decreasing the transmissions gradually tune an interferometer smoothly into a closed quantum dot. Here, we operate the device in the intermediate regime. The close relation to a quantum dot can be seen in the map of the DC current through the interferometer as a function of the applied bias and the back-gate voltage, Fig.~\ref{diamond}. This map is obtained by integrating the striped interference pattern of the differential conductance shown in Fig.~3a of the main text. ``Coulomb diamonds" can be clearly seen in the center of the map and the size of the diamonds agrees with the oscillation period $V_0$.

\begin{figure}
\centering
\includegraphics[width=1\textwidth]{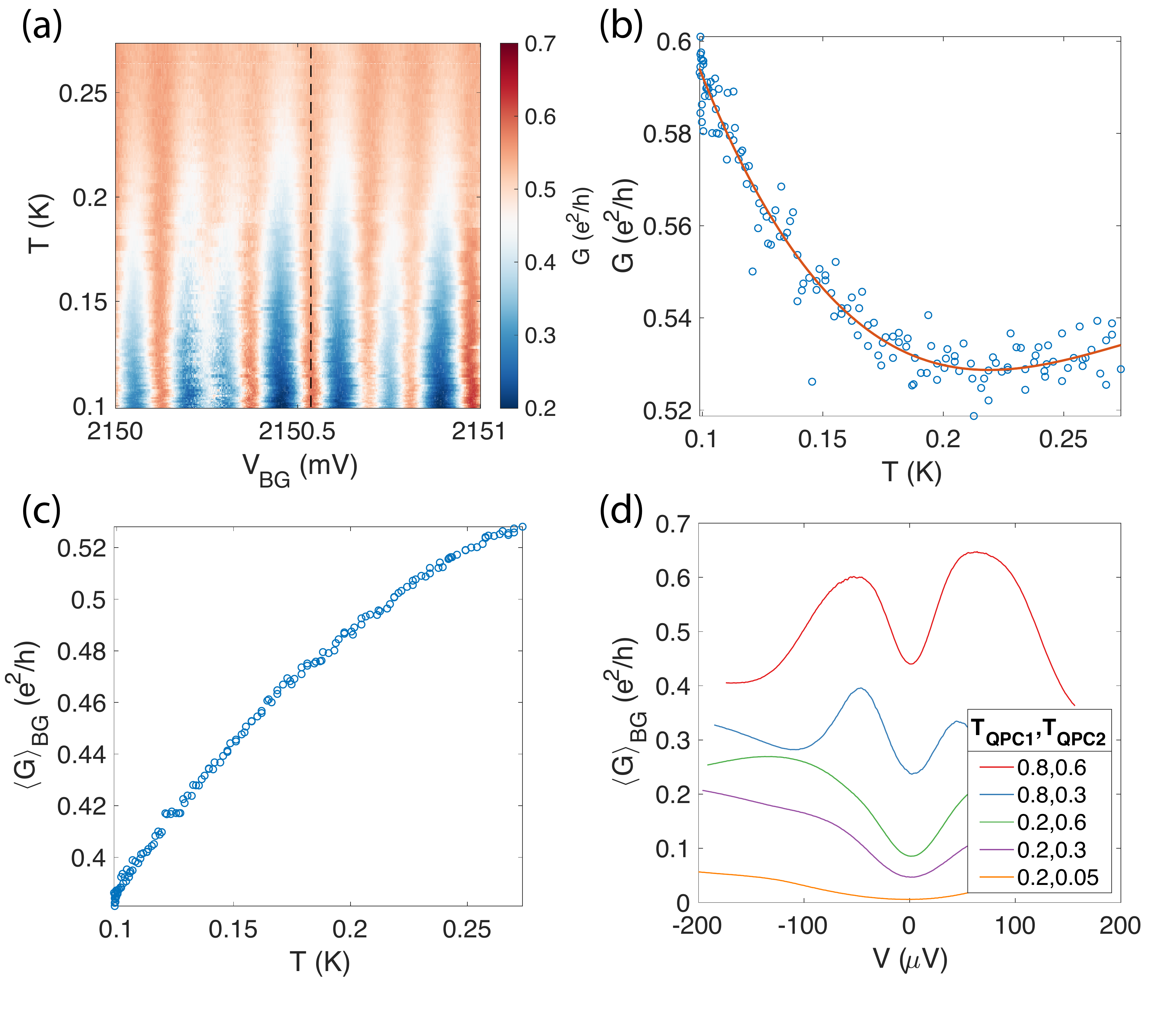}
\caption{(a) Zero-bias conductance ($\nu=1$) as a function of $V_{BG}$ and temperature, from which the visibility curve in Fig. 3(d) is extracted. (b) A cut of (a) along the dashed line showing the peak conductance as a function of temperature. The red line is a guide for the eye. (c) Zero-bias differential conductance averaged over back-gate voltage as a function of temperature. (d) Differential conductance averaged over back-gate voltage as a function of bias at $\nu=$ 1 and various QPC configurations. Here, $V_{BG}=$ 2.04 V, slightly closer to the Dirac point than the data shown in Fig.~3.}
\label{tem}
\end{figure}

\section{Interference pattern at higher energies and level spacing}
 
Fig.~\ref{tem}(a)  shows the zero-bias conductance map vs. gate voltage and temperature, measured in the same range as Fig.~3 of the main text. 
In Fig~\ref{tem}(b), we plot the peak conductance as a function of temperature following the dashed line in panel (a). Clearly, the conductance maximum, $G_{max}$, varies non-monotonically with increasing temperature. For $T<0.2$ K, $G_{max}$ decays with increasing temperature, as expected for transport through a single level. 
$G_{max}$ flattens and starts to increase above $0.2$ K, suggesting that multiple energy levels are contributing to the transport in this regime. Similarly, the contribution of the excited states is visible in the conductance averaged over gate voltage, $\langle G \rangle_{BG}$, which grows with temperature, Fig.~\ref{tem}(c). 

Finally, Fig.~\ref{tem}(d) shows $\langle G \rangle_{BG}$ measured vs. bias for several QPC configurations. In a Fabry-P\'erot interferometer with negligible charging energy, $\langle G \rangle_{BG}$ should not depend on bias (neglecting dephasing), because the same levels are contributing to transport. Both here, and in Fig.~3(a), $\langle G \rangle_{BG}$ universally demonstrates a pronounced dip around zero bias. This dip is consistent with the observed $\langle G \rangle_{BG}$ increase with temperature (Fig.~\ref{tem}(c)), thereby providing an additional indication that multiple excited levels are involved. 

We conclude that in our sample, the charging energy should dominate over the level spacing. This conclusion is consistent with the near lack of $B$ dependence in Fig.~3b.

\section{Visibility decay at finite bias}

In Fig.~3(a), we observe a rapid decay of the visibility of the interference pattern with increasing bias and the visibility becomes negligible above $\sim 100 \mu$V, close to the oscillation period $V_0$. Similar rapid decay of visibility with bias has been widely observed in both Fabry-P\'erot and Mach-Zehnder interferometers, regardless of the hosting material. Unlike Mach-Zehnder interferometers, in Fabry-P\'erot interferometers, it is hard to distinguish the effects of phase averaging and true decoherence. While the effects of decoherence can be modeled by an exponential or Gaussian decay~\cite{S_deprez_tunable_2020}, in the following we show that the observed suppression of visibility can also be phenomenologically attributed to an effective electron heating, resulting in phase averaging. 

When weakly coupled to the source and drain leads, the non-equilibrium distribution function of the edge state in the interferometer, $f_{R}$, is a weighted average of the distribution function of source and drain, $f_{S}$ and $f_{D}$~\cite{Davies1993_current}.
\begin{equation}
f_R=\frac{\Gamma_S f_S + \Gamma_D f_D}{\Gamma_S+\Gamma_D}=\gamma f_S + (1-\gamma) f_D,
\end{equation}
where $\gamma=\Gamma_S/(\Gamma_S+\Gamma_D)$ and $\Gamma_S$ ($\Gamma_D$) is the tunneling rate to the source (drain) proportional to the QPC transmissions. Strictly, this equation is only valid in the strong backscattering regime. But since it's a continuous change from strong backscattering to weak backscattering, we use it as a crude approximation for the intermediate scattering regime studied here. 
For the $\nu=1$ case presented in Fig.~3, we use $\gamma=$ 1/2 since the estimated QPC transmissions are very close to each other ($\mathcal{T}_{QPC1}\sim0.7$ and $\mathcal{T}_{QPC2}\sim0.6$).
Without energy loss to the environment, $e$-$e$ scattering tends to relax this non-equilibrium distribution into thermal equilibrium at elevated temperatures due to the excess energy in this double step distribution function~\cite{PhysRevLett.105.056803}. The effective temperature at full equilibrium is found to be~\cite{Altimiras2009}
\begin{equation}
T_{max} (V)=\sqrt{T_{SD}^2+3\gamma(1-\gamma)\left(\frac{eV}{\pi k_B}\right)^2},
\label{tmax}
\end{equation}  
where $T_{SD}$ is the electron temperature of source and drain, $V$ is the applied voltage bias and $k_B$ is the Boltzmann constant. The second term under the square root is an effective heating from the relaxation of excess energy provided by the chemical potential difference.
Since the edge channel likely does not achieve full equilibrium, we multiply this term by a parameter $\alpha<1$, which describes the amount of excess energy that has relaxed. Then, Eq.~(\ref{tmax}) becomes
\begin{equation}
T_{eff} (V,\alpha)=\sqrt{T_{SD}^2+3\gamma(1-\gamma)\left(\frac{eV}{\pi k_B}\right)^2\alpha}.
\label{teff}
\end{equation}
One can relate the relaxation parameter $\alpha$ to the electron-electron scattering rate via Boltzmann equation~\cite{Dubi2019}, which may depend on bias voltage.
As shown in the main text, the visibility, $v=(G_{max}-G_{min})/(G_{max}+G_{min})$, of AB oscillations in single-particle regime is proportional to $T/\sinh(T/T_0)$ to leading order. Therefore, at finite bias we have 
\begin{equation}
v(V)\propto \frac{T_{eff}(V,\alpha)}{\sinh(T_{eff}(V,\alpha)/T_{0})},
\label{vV}
\end{equation} 
suggesting a fast decay of visibility with increasing bias. 

\begin{figure}
\centering
\includegraphics[width=1\textwidth]{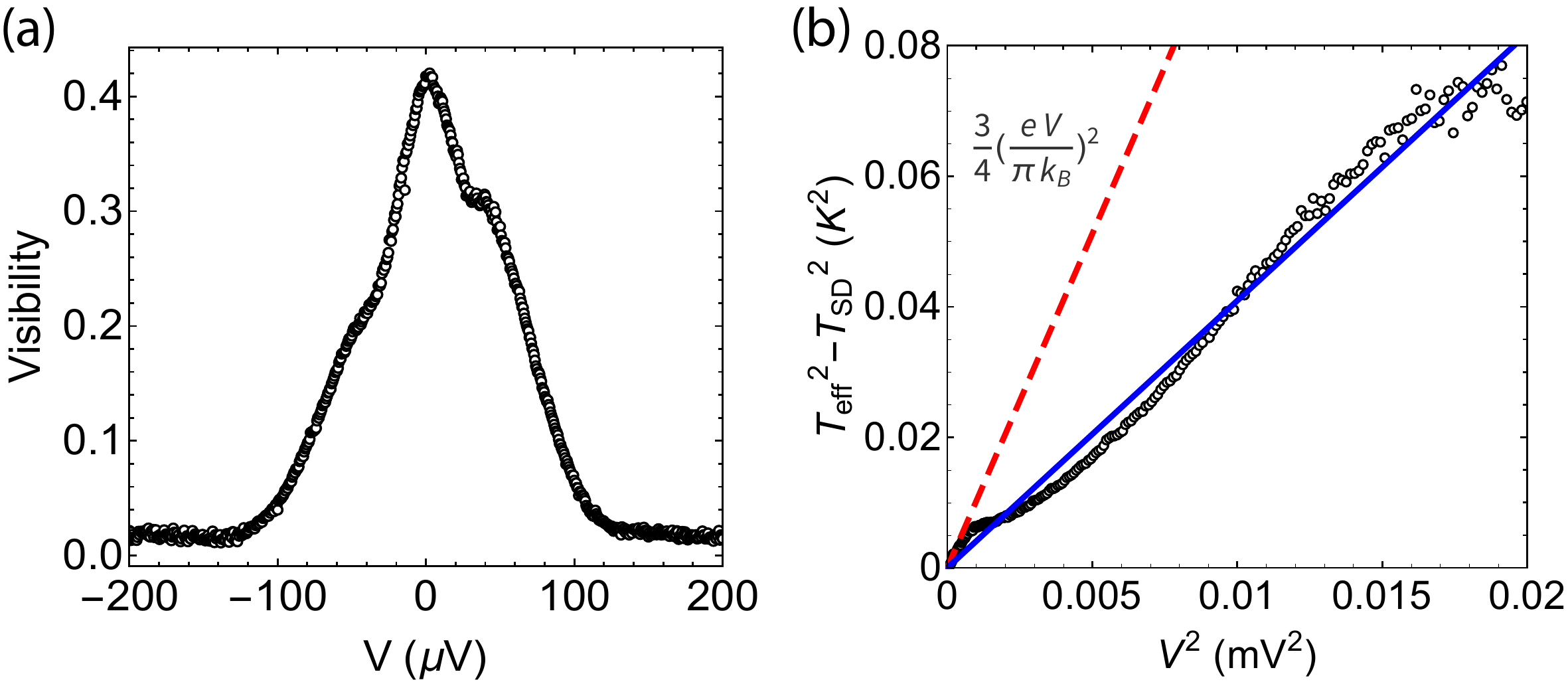}
\caption{(a) The visibility as a function of bias. The symmetrized visibility is then further converted into an effective temperature $T_{eff}$. (b) $T_{eff}^2-T_0^2$ (black dots) plotted versus $V^2$. The red dashed line plots the expectation value $\frac{3}{4}(\frac{eV}{\pi k_B})^2$ from full thermalization of a double-step distribution function with equal contribution from source and drain and the blue line plots 0.4 times this value.}
\label{visTeff}
\end{figure}

To find the dependence of $\alpha$ on $V$, we use the relation between visibility and temperature obtained in Fig.~3(d) to convert the visibility at finite bias (plotted in Fig.~\ref{visTeff} (a)) into an effective electronic temperature $T_{eff}$. Here, we symmetrize the visibility measured at positive and negative bias since the visibility decay is mostly independent from the small bias asymmetry observed in Fig.~\ref{visTeff}(a).
In Fig.~\ref{visTeff}(b), we plot $T_{eff}^2-T_{SD}^2$ inferred from the visibility as a function of $V^2$ (black dots), together with $T_{max}^2-T_{SD}^2$ from Eq.~(\ref{tmax}) (red dashed line). First, we see that $T_{eff}$ does not exceed $T_{max}$, confirming that attributing the visibility drop to effective electron equilibration is not unreasonable. Second, $T_{eff}^2-T_{SD}^2$ is close to about 0.4 times $T_{max}^2-T_{SD}^2$ (blue line), indicating that about 40$\%$ of the excess energy provided by the chemical potential difference is converted into thermal energy. Surprisingly, this fraction does not strongly depends on bias. 
Therefore, we find a bias independent $\alpha=0.4$ that can well explain the finite bias visibility decay and from now on we treat it as a known constant for simulating the conductance map shown in Fig.~3(a). 

Historically, the bias dependence of visibility of interference patterns has been fitted to either exponential decay~\cite{mcclure_edge-state_2009} for Fabry-P\'erot interferomters or Gaussian decay~\cite{Roulleau_finitebias_2007} for Mach-Zehnder interferometers. A comparison of both fits in Fabry-P\'erot interferomters has been presented in Ref.~\cite{S_deprez_tunable_2020}, where both fitting expressions can capture the fast decay but not the exact trend. Here, Eq.~(\ref{teff}\&\ref{vV}) produces a Gaussian decay at low bias ($eV<k_BT_{SD}$) and an exponential decay at high bias in agreement with the refined Gaussian phase randomization model presented in Ref.~\cite{Roulleau2008}.

\begin{figure}
\centering
\includegraphics[width=1\textwidth]{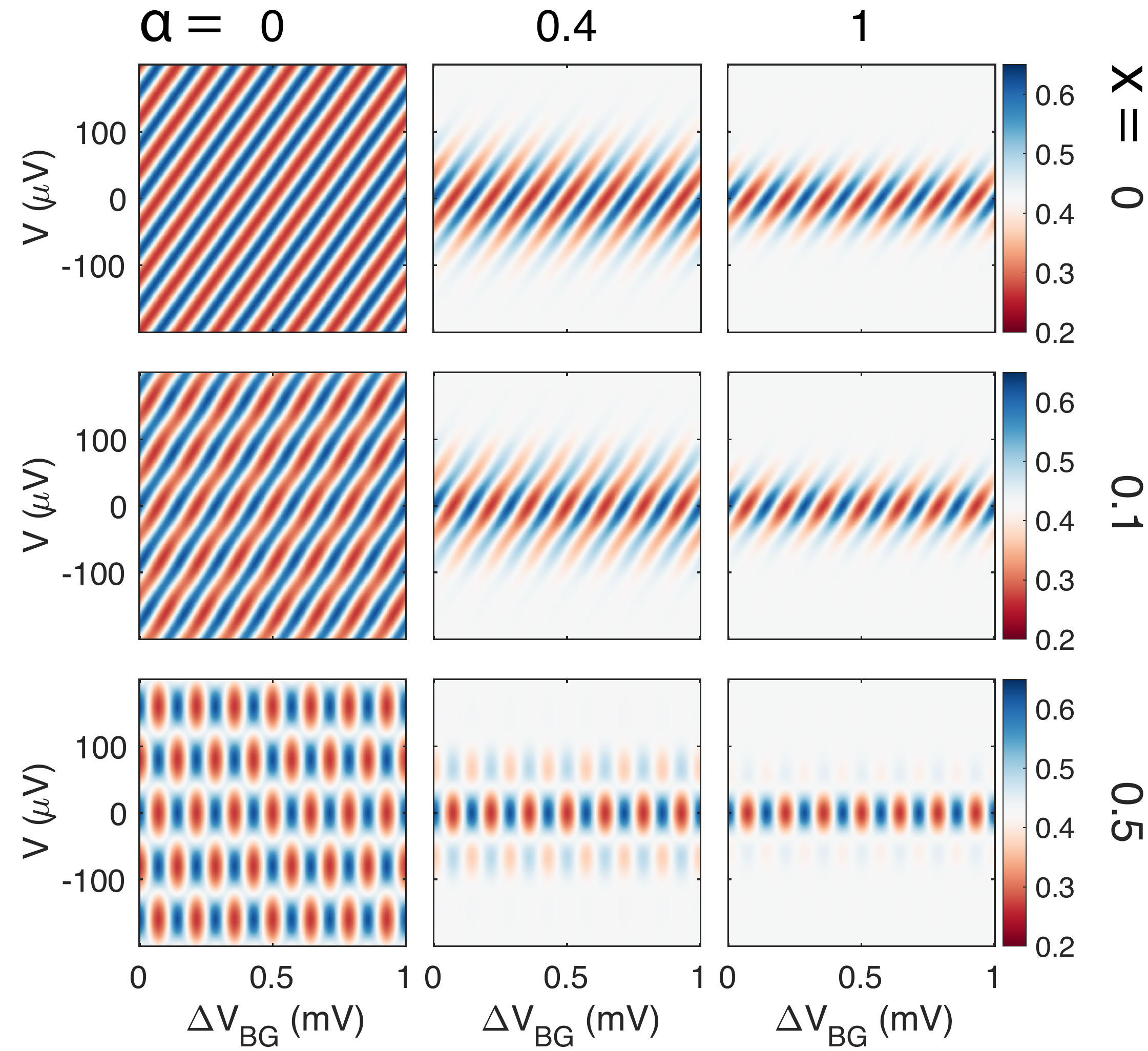}
\caption{Calculated conductance maps as a function of bias and back-gate voltage for various $x$ and $\alpha$. From top to bottom, $x=$ 0, 0.1 and 0.5. From left to right, $\alpha=$ 0, 0.4 and 1. The unit of the colorbar is $e^2/h$.}
\label{9map}
\end{figure}

Next, instead of approximating the interference pattern as a sinusoidal function, we also want to take into account higher order terms and the bias symmetrization effect observed widely in prior works.
Neglecting the renormalization of QPC transmissions which can give a power law modulation~\cite{NgoDinh2012}, the total current through the interferometer at zero temperature is 
\begin{equation}
I=\frac{-e}{h}\int_{-eV_D}^{-eV_S}\frac{\mathcal{T}_{QPC1}\mathcal{T}_{QPC2}}{1+\mathcal{R}_{QPC1}\mathcal{R}_{QPC2}-2\sqrt{\mathcal{R}_{QPC1}\mathcal{R}_{QPC2}}\cos(\delta\phi+\frac{2\pi E}{eV_0})}dE,
\end{equation}
where $\mathcal{R}_{QPC1}=1-\mathcal{T}_{QPC1}$ and $\mathcal{R}_{QPC2}=1-\mathcal{T}_{QPC2}$. The distribution of the voltage drop is determined by the capacitive coupling between the interferometer and the source, drain, and gate electrodes. To take this effect into account, we follow Ref.~\cite{S_deprez_tunable_2020} to set the source and drain bias drops as $V_{S}=(1-x)V$ and $V_{D}=-xV$.
Therefore, the zero-temperature differential conductance $G_0=dI/dV$ at finite bias $V$ is 
\begin{equation}
G_0(V,x)=(1-x)g_0(-eV_S)+xg_0(-eV_D),
\end{equation}
where
\begin{equation}
g_0(E)=\frac{e^2}{h}\frac{\mathcal{T}_{QPC1}\mathcal{T}_{QPC2}}{1+\mathcal{R}_{QPC1}\mathcal{R}_{QPC2}-2\sqrt{\mathcal{R}_{QPC1}\mathcal{R}_{QPC2}}\cos(\delta\phi+\frac{2\pi E}{eV_0})}
\end{equation}
is the differential conductance contribution from electrons at energy $E$.

The differential conductance at $T_{eff} (V,f)$ is then 
\begin{equation}
G(V,\alpha,x)=(1-x)g(-eV_S,V,\alpha)+xg(-eV_D,V,\alpha),
\end{equation}
where
\begin{equation}
g(E,V,\alpha)=\int_{-\infty}^{\infty} g_0 (E+\epsilon) \frac{\partial}{\partial \epsilon}\left(\frac{1}{e^{\epsilon/k_BT_{eff} (V,\alpha)}+1}\right) d\epsilon
\end{equation}

\begin{figure}
\centering
\includegraphics[width=0.6\textwidth]{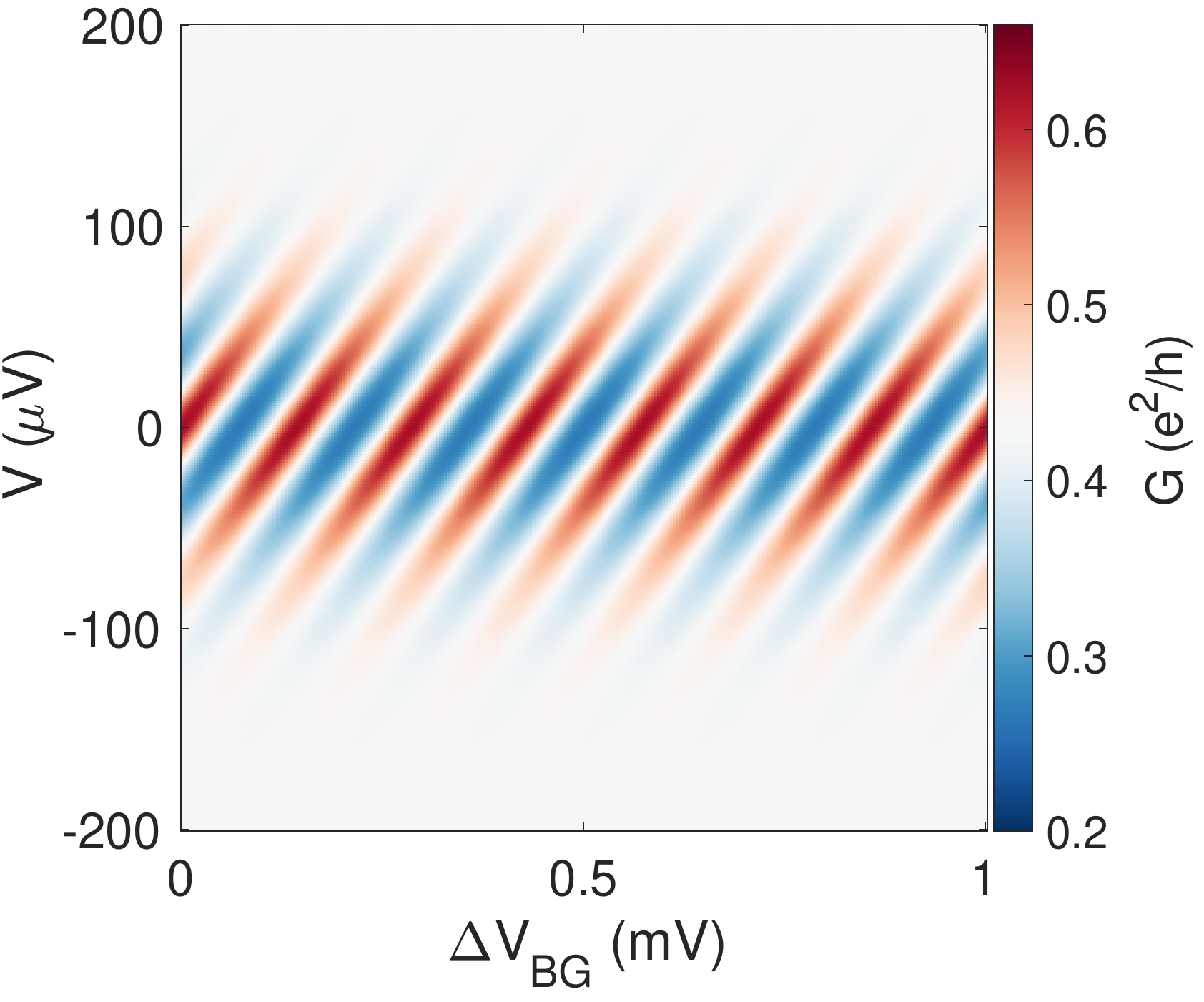}
\caption{Calculated conductance maps as a function of bias and back-gate voltage for to simulate the experimental result in Fig.~3(c).}
\label{simmap}
\end{figure}

To demonstrate the bias symmetrization effect together with the electron heating effect, we plot the calculated conductance maps for $x=$ 0, 0.1, 0.5 and $\alpha=$ 0, 0.4, 1 in Fig.~\ref{9map}. As the voltage drop on drain increases, the pattern gradually evolves into a checkerboard. Meanwhile, increasing electron thermalization results in faster visibility decay at high bias. To simulate the background subtracted map shown in Fig.~3(c), in Fig.~\ref{simmap}, we use $\mathcal{T}_{QPC1}=\mathcal{T}_{QPC2}=0.6$, $T_0=$ 0.1 K, $\alpha=$ 0.4 and $x=$ 0. The map clearly reproduces the rapid decay of visibility above $\sim 100 \mu$V seen in Figs.~3(a,c). Since there is no voltage drop on drain ($x=$ 0), the obtained interference pattern is stripe in contrast to the widely observed checkerboard pattern in GaAs. The checkerboard pattern corresponds to a symmetric voltage drop on source and drain due to the electrochemical potential of the cavity being tied to the source-drain voltages~\cite{NgoDinh2012}. Ref.~\cite{S_deprez_tunable_2020} also observed a small voltage drop at the drain in a graphene interferometer of similar size and the effect of symmetrization increases with the interferometer size.

\providecommand{\latin}[1]{#1}
\providecommand*\mcitethebibliography{\thebibliography}
\csname @ifundefined\endcsname{endmcitethebibliography}
  {\let\endmcitethebibliography\endthebibliography}{}

\end{document}